\documentclass[aps, pra, 10pt, superscriptaddress, twocolumn]{revtex4-1}
\usepackage{stmaryrd}
\usepackage{amsfonts}
\usepackage{mathrsfs}
\usepackage{amsmath}
\usepackage{amscd}
\usepackage{graphicx}

\begin{document}
\title{Experimental Realization of Single-shot Nonadiabatic Holonomic Gates in Nuclear Spins}

\author{Hang Li}
\affiliation{State Key Laboratory of Low-Dimensional Quantum Physics and Department of Physics, Tsinghua University, Beijing 100084, China}
\affiliation{Collaborative Innovation Centre of Quantum Matter, Beijing 100084, China}
\author{Guilu Long}
\email{gllong@mail.tsinghua.edu.cn}
\affiliation{State Key Laboratory of Low-Dimensional Quantum Physics and Department of Physics, Tsinghua University, Beijing 100084, China}
\affiliation{Collaborative Innovation Centre of Quantum Matter, Beijing 100084, China}
\begin{abstract}
Nonadiabatic holonomic quantum computation has received increasing attention due to its robustness against control errors. However, all the previous schemes have to use at least two sequentially implemented gates to realize a general one-qubit gate. Based on two recent works~\cite{xu2015nonadiabatic,sjoqvist2016nonadiabatic}, we construct two Hamiltonians and experimentally realized nonadiabatic holonomic gates by a single-shot implementation in a two-qubit nuclear magnetic resonance (NMR) system. Two noncommuting one-qubit holonomic gates, rotating along $\hat{x}$ and $\hat{z}$ axes respectively, are implemented by evolving a work qubit and an ancillary qubit nonadiabatically following a quantum circuit designed. Using a sequence compiler developed for NMR quantum information processor, we optimize the whole pulse sequence, minimizing the total error of the implementation. Finally, all the nonadiabatic holonomic gates reach high unattenuated experimental fidelities over $98\%$.

\end{abstract}
\maketitle

\section{Introduction} 
Quantum computation provides an unprecedented computaional power over classical computation. With the quantum parallelism, quantum algorithms such as Shor's factoring algorithm~\cite{shor1994algorithms} and Gover's searching algorithm~\cite{grover1997quantum}, provide strong evidences that quantum computation can gain exponential speed-up in many practical problems. However, on the way to realizing practical scalable quantum information processing, errors from quantum gates in the control process and decoherence caused by the inevitable interaction between the physical system and its environment are two main difficulties encountered.

Adiabatic holonomic quantum computation (AHQC), as a promising quatum computation model, was first proposed by Zanardi and Rasetti~\cite{zanardi1999holonomic} based on Wilczek-Zee geometric phases. They discovered that non-Abelian geometric phase can be used for implementing robust quantum gates, which is fault tolerant with certain errors in the control process, by encoding quantum information into degenerate energe subspaces of a Hamiltonian depending on parameters and adiabatically evoluting the quantum states along a loop in the corresponding parameter space. Further more, many adiabatic holonomic gates schemes have been proposed for trapped ions~\cite{duan2001geometric}, superconducting qubits~\cite{faoro2003non} and semiconductor quantum dots~\cite{solinas2003semiconductor}.

Due to the long evolution time needed to fulfill the adiabatic condition, AHQC is diffcult to realize experimentally. Nonadiabatic holonomic quantum computation (NHQC)~\cite{sjoqvist2012non,xu2012nonadiabatic} was proposed to avoid the long run-time requirement based on nonadiabatic and non-Abelian geometric phases~\cite{anandan1988non}, while retaining the advantages of geometric nature. Although NHQC was just proposed recently, it gains increasing attentions because of its combining speed and universality, where lots of theoretical schemes~\cite{johansson2012robustness,spiegelberg2013validity,mousolou2014universal,mousolou2014nonabelian,zhang2014quantum,xu2014protecting,xu2014universal,liang2014nonadiabatic,zhang2014fast,
zhou2015cavity,xue2015universal,sjoqvist2015geometric,pyshkin2015expedited,xu2015nonadiabatic,sjoqvist2016nonadiabatic,song2016shortcuts} are prompted and meanwhile NHQC has been realized experimentally in NMR~\cite{feng2013experimental}, superconducting quantum devices~\cite{abdumalikov2013experimental} and diamond nitrogen-vacancy centers~\cite{zu2014experimental,arroyo2014room} recently.

All the previous shemes of NHQC have to use at least two sequentially implemented gates to realize a general one-qubit gate. Two recent schemes~\cite{xu2015nonadiabatic,sjoqvist2016nonadiabatic} for one-qubit nonadiabatic holonomic gates were put forward where one can avoid the extra work of combining two gates into one, realizing an arbitrary one-qubit gate with a single-shot implementation. In this letter, we report the first experimental realization of one-qubit nonadiabatic holonomic gate of single-shot scheme in a liquid NMR quantum information processor. By modifying the original three-level quantum system scheme, a two-qubit quantum system instead is constructed with a work qubit and an ancillary qubit. To prove the realization of single-shot holonomic gates, nonadiabatic holonomic rotations along $\hat{x}$ and $\hat{z}$ axes respectively as examples are implemented experimentally in a two-qubit NMR quantum information processor.

\section{The single-shot scheme} 
In Ref.~\cite{xu2015nonadiabatic,sjoqvist2016nonadiabatic}, consider a three-level quantum system with driving laser fields, and its effective Hamiltonian
\begin{equation}
\begin{aligned}
H_{eff}=&\Omega\sin\gamma(|e\rangle\langle e|+|b\rangle\langle b|)+\Omega[\cos\gamma(|b\rangle\langle e|\\&+|e\rangle\langle b|)+\sin\gamma(|e\rangle\langle e|-|b\rangle\langle b|)],
\end{aligned}
\end{equation}
where $|b\rangle=\cos\alpha|0\rangle+e^{i\beta}\sin\alpha|1\rangle$ and $|d\rangle=\sin\alpha|0\rangle-e^{i\beta}\cos\alpha|1\rangle$. After a time period $T=\pi/\Omega$, evolution operator $U(T)$ in the basis $\{|e\rangle, |b\rangle, |d\rangle\}$ can be written as
\begin{equation}
U(T)=e^{-iH_{eff}T}=\left(
\begin{matrix}
e^{-i\phi}&0&0\\0&e^{-i\phi}&0\\0&0&1
\end{matrix}\right),
\end{equation}
where $\phi=\pi\sin\gamma+\pi$. $U(T)$ is equivalent to $U^L(T)$ which is in the logical subspace spanned by $\{|0\rangle, |1\rangle\}$
\begin{equation}
U^L(T)=e^{-i\frac{\phi}{2}(|b\rangle\langle b|-|d\rangle\langle d|)}.
\end{equation}
By selecting proper parameters $\alpha$, $\beta$ and $\gamma$, one can achieve arbitrary one-qubit holonomic gate with single-shot implementation.

Now, let's review the holonomic conditions. Consider an $N$-dimensional quantum system with its Hamiltonian $H(t)$. Assume the state of the system is initially in a $M$-dimensional subspace $\mathcal{S}(0)$ spanned by a set of orthonormal basis vectors $\left\{\left|\phi_k(0)\right\rangle\right\}^M_{k=1}$ . It has been proven that~\cite{sjoqvist2012non,xu2012nonadiabatic} the evolution operator is a holonomic matrix acting on $\mathcal{S}(0)$ if $\left|\phi_k(t)\right\rangle$ satisfy the following conditions:
\begin{equation}
\left(i\right)\ \sum^M_{k=1}\left|\phi_k(\tau)\right\rangle\left\langle\phi_k(\tau)\right|=\sum^M_{k=1}\left|\phi_k(0)\right\rangle\left\langle\phi_k(0)\right|
\end{equation}
\begin{equation}
\left(ii\right)\ \left\langle\phi_k(t)\right|H(t)\left|\phi_l(t)\right\rangle=0,\quad k,l=1, ..., M
\end{equation}
where $\tau$ is the evolution period and $\left|\phi_k(t)\right\rangle=\mathbf{T} \exp[-i\boldsymbol{\int}^t_0H(t_1)dt_1]\left|\phi_k(0)\right\rangle$ and $\mathbf{T}$ being time odering.
 
Here we construct a four-level system consisting of two qubits and label a subspace $\mathcal{S}^L=\{\left|10\right\rangle, \left|11\right\rangle\}$ as our logical working space, i.e., $\left|0\right\rangle_{L}\equiv\left|10\right\rangle$, $\left|1\right\rangle_{L}\equiv\left|11\right\rangle$. In such way, the first qubit acts as an ancillary qubit and the whole information of the logical qubit is encoded in the work qubit (the second qubit). To realize two noncommuting nonadiabatic one-qubit gates, simply, two rotation operators along $\hat{x}$ axis and $\hat{z}$ axis respectively, $R_{x}^{L}(\theta)$ and $R_{z}^{L}(\theta)$, two different Hamiltonians $H_x$ and $H_z$ are designed to match their respective rotations, as presented below:
\begin{equation}
H_x=H_1+H_2,\ H_z=H_1+H_3,
\end{equation}
\begin{equation}
H_1=\frac{\Omega\sin\lambda}{2}\left(I+\sigma^1_z-\sigma^2_z-\sigma^1_z\sigma^2_z\right),
\end{equation}
\begin{equation}
H_2=\frac{\Omega\cos\lambda}{2\sqrt{2}}\left(\sigma^1_x\sigma^2_x+\sigma^1_y\sigma^2_y+\sigma^1_x-\sigma^1_x\sigma^2_z\right),
\end{equation}
\begin{equation}
H_3=\frac{\Omega\cos\lambda}{2}\left(\sigma^1_x\sigma^2_x+\sigma^1_y\sigma^2_y\right),
\end{equation}
where $I$ is a $4\times4$ dimensional identity matrix, and $\sigma_x$, $\sigma_y$ and $\sigma_z$ are Pauli matrices. In another way, expressing $H_x$ and $H_z$ as matrices in the basis $\{\left|00\right\rangle, \left|01\right\rangle, \left|10\right\rangle, \left|11\right\rangle\}$:
\begin{equation}\label{e1}
H_x=\Omega\left(
\begin{matrix}
0&0&0&0\\
0&2\sin\lambda&\frac{\cos\lambda}{\sqrt{2}}&\frac{\cos\lambda}{\sqrt{2}}\\
0&\frac{\cos\lambda}{\sqrt{2}}&0&0\\
0&\frac{\cos\lambda}{\sqrt{2}}&0&0
\end{matrix}
\right),
\end{equation}
\begin{equation}\label{e2}
H_z=\Omega\left(
\begin{matrix}
0&0&0&0\\
0&2\sin\lambda&\cos\lambda&0\\
0&\cos\lambda&0&0\\
0&0&0&0
\end{matrix}
\right).
\end{equation}

After a proper time $\tau=\pi/\Omega$, the two noncommuting rotation operations $R_x^L(\theta)$ and $R_z^L(\theta)$ generated by their respective Hamiltonians $H_x$ and $H_z$ read
\begin{equation}\label{e3}
R_x^L(\theta)=e^{-i\dfrac{\theta}{2}}\left(
\begin{matrix}
e^{i\frac{\theta}{2}}&0&0&0\\
0&e^{-i\frac{\theta}{2}}&0&0\\
0&0&\cos{\frac{\theta}{2}}&-i\sin{\frac{\theta}{2}}\\
0&0&-i\sin{\frac{\theta}{2}}&\cos{\frac{\theta}{2}}
\end{matrix}
\right),
\end{equation}
\begin{equation}\label{e4}
R_z^L(\theta)=e^{-i\frac{\theta}{2}}\left(
\begin{matrix}
e^{i\frac{\theta}{2}}&0&0&0\\
0&e^{-i\frac{\theta}{2}}&0&0\\
0&0&e^{-i\frac{\theta}{2}}&0\\
0&0&0&e^{i\frac{\theta}{2}}
\end{matrix}
\right),
\end{equation}
where $\theta=\pi\left(\sin{\lambda}+1\right)$. According to Eqs. (\ref{e1})-(\ref{e4}) and subspace $\mathcal{S}^L$, we can easily prove both holonomic conditions (i) and (ii) are satisfied, thus, both rotation operations $R_x^L(\theta)$ and $R_z^L(\theta)$ generated by $H_x$ and $H_z$ respectively are one-qubit holonomic gates acting on subspace $\mathcal{S}^L$. In the logical subspace $\mathcal{S}^L$ spanned by $\left|10\right\rangle$ and $\left|11\right\rangle$, $R_x^{L}(\theta)$ and $R_z^{L}(\theta)$ can be expressed as
\begin{equation}
R_x^L(\theta)\sim\left(
\begin{matrix}
\cos{\frac{\theta}{2}}&-i\sin{\frac{\theta}{2}}\\
-i\sin{\frac{\theta}{2}}&\cos{\frac{\theta}{2}}
\end{matrix}
\right),
\end{equation}
\begin{equation}
R_z^L(\theta)\sim\left(
\begin{matrix}
e^{-i\frac{\theta}{2}}&0\\
0&e^{i\frac{\theta}{2}}
\end{matrix}
\right),
\end{equation}
from which we can see $R_x^L(\theta)$ and $R_z^L(\theta)$ are exactly one-qubit rotation operations along $x$ and $z$ axis, respectively. Till now, we construct two one-qubit noncommuting nonadiabatic holonomic gates and an arbitrary one-qubit nonadiabatic holonomic gate can be constructed in the same way with a single-shot implementation.

Since Hamiltonians $H_x$ and $H_z$ are both time-independent, $R_x^L(\theta)$ and $R_z^L(\theta)$ can be split into $N$ slices with equal time interval, like
\begin{equation}
R_x^L(\theta)=\prod\nolimits_{k=1}^Ne^{-iH_x\tau/N},
\end{equation}
\begin{equation}
R_z^L(\theta)=\prod\nolimits_{k=1}^Ne^{-iH_z\tau/N}.
\end{equation}
If $N$ is large enough, in other words, each time interval $\Delta\tau\equiv\tau/N$ is small enough, both $R_x^L(\theta)$ and $R_z^L(\theta)$ evolutions in each time interval can be approximated into a sequence of noncommuting evolutions in the order of $\left(\Delta\tau\right)^2$ by the aid of the Trotter formula~\cite{nielsen2010quantum}. Here we take $R_x^L(\theta)$ evolution in a small time interval $\Delta\tau$ as an example,
\begin{equation}
R_x^L(\theta/N)\approx e^{-i(\Delta\tau/2)H_1}e^{-i\Delta\tau H_2}e^{-i(\Delta\tau/2)H_1},
\end{equation}
as each component of $H_1$ commutes with each other, here we focus on expanding $e^{-i\Delta\tau H_2}$, defining $\varphi\equiv\Delta\tau\Omega\cos\lambda/(2\sqrt{2})=\pi\cos\lambda/(2\sqrt{2}N)$,
\begin{equation}
e^{-i\Delta\tau H_2}=e^{-i\varphi(\sigma_x^1\sigma_x^2-\sigma_x^1\sigma_z^2)}e^{-i\varphi(\sigma_y^1\sigma_y^2+\sigma_x^1)},
\end{equation}
as a fact of $\sigma_x^1\sigma_x^2-\sigma_x^1\sigma_z^2$ commuting with $\sigma_y^1\sigma_y^2+\sigma_x^1$. Utilizing a method of constructing a new angular momentum vector in Ref.~\cite{zhang2005simulation}, two components of $e^{-i\Delta\tau H_2}$ denote:
\begin{equation}
e^{-i\varphi(\sigma_x^1\sigma_x^2-\sigma_x^1\sigma_z^2)}=e^{-i\frac{\pi}{8}\sigma_y^2}e^{-i\frac{\pi\cos\lambda}{2N}\sigma_x^1\sigma_x^2}e^{i\frac{\pi}{8}\sigma_y^2},
\end{equation}
\begin{equation}
e^{-i\varphi(\sigma_y^1\sigma_y^2+\sigma_x^1)}=e^{-i\frac{\pi}{8}\sigma_z^1\sigma_y^2}e^{-i\frac{\pi\cos\lambda}{2N}\sigma_x^1}e^{i\frac{\pi}{8}\sigma_z^1\sigma_y^2}.
\end{equation}

In conclusion, $R_x^L(\theta)$ can be approximately realized by a sequence of one-qubit rotations and evolutions of $J$-coupling constant in a two-qubit NMR system:
\begin{equation}
	\begin{aligned}
&R_x^L(\theta)\approx\prod\nolimits_{k=1}^N(e^{-i\frac{\pi\sin\lambda}{4N}\sigma_z^1}e^{i\frac{\pi\sin\lambda}{4N}\sigma_z^2}e^{i\frac{\pi\sin\lambda}{4N}\sigma_z^1\sigma_z^2}\\
&\times e^{-i\frac{\pi}{8}\sigma_z^1\sigma_y^2}e^{-i\frac{\pi\cos\lambda}{2N}\sigma_x^1}e^{i\frac{\pi}{8}\sigma_z^1\sigma_y^2}e^{-i\frac{\pi}{8}\sigma_z^1\sigma_y^2}e^{-i\frac{\pi\cos\lambda}{2N}\sigma_x^1}\\
&\times e^{i\frac{\pi}{8}\sigma_z^1\sigma_y^2}e^{-i\frac{\pi\sin\lambda}{4N}\sigma_z^1}e^{i\frac{\pi\sin\lambda}{4N}\sigma_z^2}e^{i\frac{\pi\sin\lambda}{4N}\sigma_z^1\sigma_z^2}).
	\end{aligned}
\end{equation}
Simultaneously, $R_z^L(\theta)$ can be approximated as
\begin{equation}
	\begin{aligned}
		R_z^L(\theta)\approx&\prod\nolimits_{k=1}^N(e^{-i\frac{\pi\sin\lambda}{4N}\sigma_z^1}e^{i\frac{\pi\sin\lambda}{4N}\sigma_z^2}e^{i\frac{\pi\sin\lambda}{4N}\sigma_z^1\sigma_z^2}\\
		&\times e^{-i\frac{\pi\cos\lambda}{4N}\sigma_x^1\sigma_x^2}e^{-i\frac{\pi\cos\lambda}{4N}\sigma_y^1\sigma_y^2}e^{-i\frac{\pi\sin\lambda}{4N}\sigma_z^1}\\
		&\times e^{i\frac{\pi\sin\lambda}{4N}\sigma_z^2}e^{i\frac{\pi\sin\lambda}{4N}\sigma_z^1\sigma_z^2}).
	\end{aligned}
\end{equation}
\begin{figure}[!t]
\begin{center}
\includegraphics[width=8.5cm,angle=0]{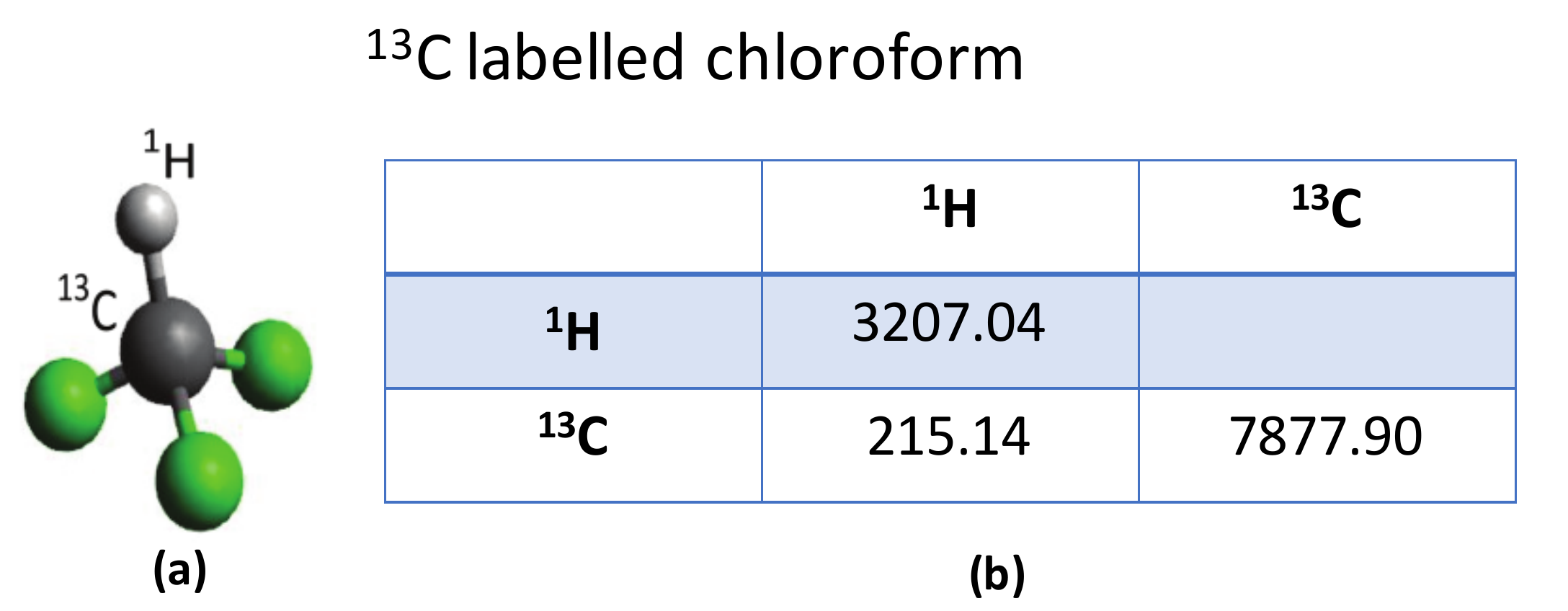}
\caption{Molecular structure (a) and Hamiltonian parameters (b) of chloroform. The chemical shifts and scalar coupling constant of the molecule are on and below the diagonal (in Hz) in the table of (b), respectively.}\label{f1}
\end{center}
\end{figure}
\paragraph*{\bf Experimental background.} 
Nuclear magnetic resonance (NMR) is a reliable technology for studying small-to-medium size quantum information experiments~ \cite{peng2015zeros,lu2015clifford,zhang2012magnet}, and quantum simulation~\cite{jin2016experimental,Lu2015,liu2015first,pearson2016experimental}. Motivated by the needs of studying quantum information, many sophisticated techniques of controlling nuclear spins have been developed. 

In this paper, all the experiments are carried out at room temperature (295 K) on a Bruker Avance III 400 MHz spectrometer and we used the $^{13}$C labelled chloroform dissolved in $d6$ acetone as a two-qubit NMR quantum information processor. The structure and Hamiltonian parameters of chloroform are shown in Fig.~\ref{f1}, where $^1$H and $^{13}$C nuclear spins respectively act as an ancillary qubit and our work qubit. Moreover, the internal Hamiltonian of the system is given by
\begin{equation}
H_{int}=\omega_1\sigma^{1}_{z}+\omega_2\sigma^{2}_{z}+{\frac{\pi}{2}J\sigma^1_z\sigma^2_z}.
\end{equation}
\begin{figure*}
\begin{center}
\includegraphics[width=15cm, angle=0]{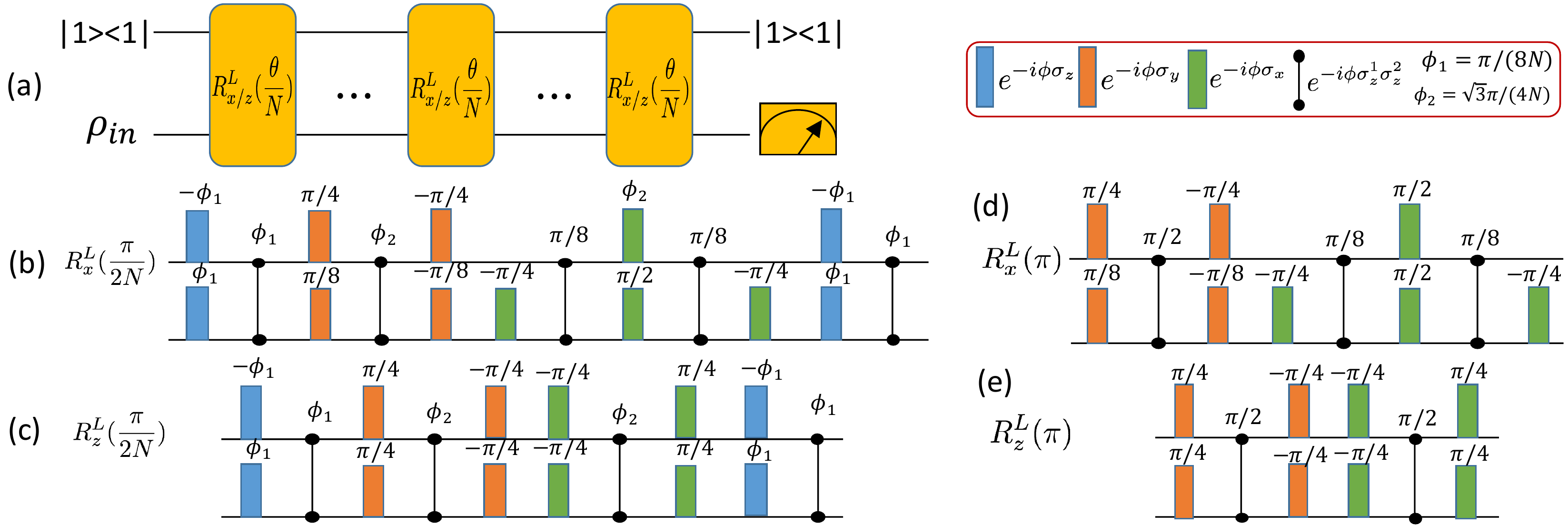}
\caption{(a) Quantum circuit for implementing one-qubit NHQC gates, where the ancillary qubit stays at $|1\rangle\langle1|$ before and after the circuit and only the work qubit is measured in the final stage. (b) and (c) are approximated pulse sequences for implementing one-qubit NHQC rotations $R_x^{L}(\frac{\pi}{2N})$ and $R_z^{L}(\frac{\pi}{2N})$ based on the Trotter formula respectively, while (d) and (e) are the exact pulse sequences for $R_x^{L}(\pi)$ and $R_z^{L}(\pi)$ respectively. }\label{f2}
\end{center}
\end{figure*}
\section{Experimental details} 
The whole experimental procedure consists of three parts: (i) quantum state preparation, (ii) implementation of two noncommuting nonadiabatic one-qubit holonomic gates, and (iii) quantum process characterization of NHQC gates.

Our logical subspace locates at $\{\left|10\right\rangle, \left|11\right\rangle\}$ and holonomic conditions require the NMR system ends in the logical subspace after applying NHQC gates as it starts in the logical subspace. At the quantum state preparation stage, the ancillary qubit $^{1}$H is prepared in state $\left|1\rangle\langle1\right|$ and the work qubit $^{13}$C is prepared in $\rho_{in}$. Writing $\rho_{in}$ in form of deviation matrix~\cite{chuang1998bulk}, here we can get a set of $\rho_{in}$:
\begin{equation}\label{e5}
\rho_{in}\in\{\sigma_x, \sigma_y, \sigma_z\}.
\end{equation}
Hence the NMR system is prepared in state $\left|1\rangle\langle1\right|\otimes\rho_{in}$, which is realized by a cat-state method~\cite{knill2000algorithmic,souza2011experimental}.

At the stage of implementation of one-qubit holonomic gates, we selected four one-qubit NHQC gates, $R_x^{L}(\pi/2)$, $R_z^{L}(\pi/2)$, $R_x^{L}(\pi)$ and $R_z^{L}(\pi)$, and then decomposed them into a combination of radio-frequency pulses and evolutions of $J$-coupling constants between the two qubits that can be manipulated directly in the NMR system~\cite{vandersypen2005nmr,cory2000nmr,jones2011quantum}. Fig.~\ref{f2}(a) illustrates a quantum circuit for implementing a ratation $R_{x/z}^{L}(\theta)$ with an arbitrary angle $\theta$, while approximated pulse sequences for implementing $R_x^{L}(\pi/(2N))$ and $R_z^{L}(\pi/(2N))$ based on the Trotter formula are presented in Fig.~\ref{f2}(b) and (c) respectively. To be noticed, $R_x^{L}(\pi)$ and $R_z^{L}(\pi)$ can be implemented in NMR pulse sequences directly without approximatations, and their pulse sequences are shown in Fig.~\ref{f2}(d) and (e) respectively. In both case of implementing $R_x^{L}(\pi/2)$ and $R_z^{L}(\pi/2)$, we take $N=3$ and the errors between them are both below $1\%$. Among the above pulse sequences, $e^{-i\phi\sigma_z}$ in Fig.~\ref{f2}(b) - (e) is not a hard pulse that can apply to the qubit directly, but a phase evolution and implemented by simply modifying the phases of subsequent pulses and potentially the observation phase. In running the quantum circuit, a sequence compiler developed for NMR quantum information processor based on Ref.~\cite{ryan2008liquid} is used here to optimize the whole pulse sequence, minimizing the total error of the implementation compared to the intended evolution.

A method for a usual full characterization of a quantum process is quantum process tomography (QPT)~\cite{chuang1997prescription,poyatos1997complete,Wang2016}, which can quantitatively describe the implementation of one-qubit NHQC gates. By the definition of QPT, an input state $\rho_{in}$ becomes its output state $\rho_{out}$ through a quantum channel, and $\rho_{out}$ can be written as
\begin{equation}
\rho_{out}=\sum\nolimits_{k,l}\chi_{k,l}e_k\rho_{in}e_l^{\dagger},
\end{equation}
where $e_k$ is a basis for operators on the state space which forms a fixed basis operator set. As a matter of convenience, we choose $e_k$ as
\begin{equation}
e_k\in\{I, \sigma_x, -i\sigma_y, \sigma_z\},\quad k=1, ..., 4.
\end{equation}
Thus, our one-qubit NHQC gates can be characterized by a matrix $\chi$. By determining experimental $\chi$, we prove a good implementation of one-qubit NHQC gates.
\begin{figure}[!h]
\begin{center}
\includegraphics[width=8.5cm,angle=0]{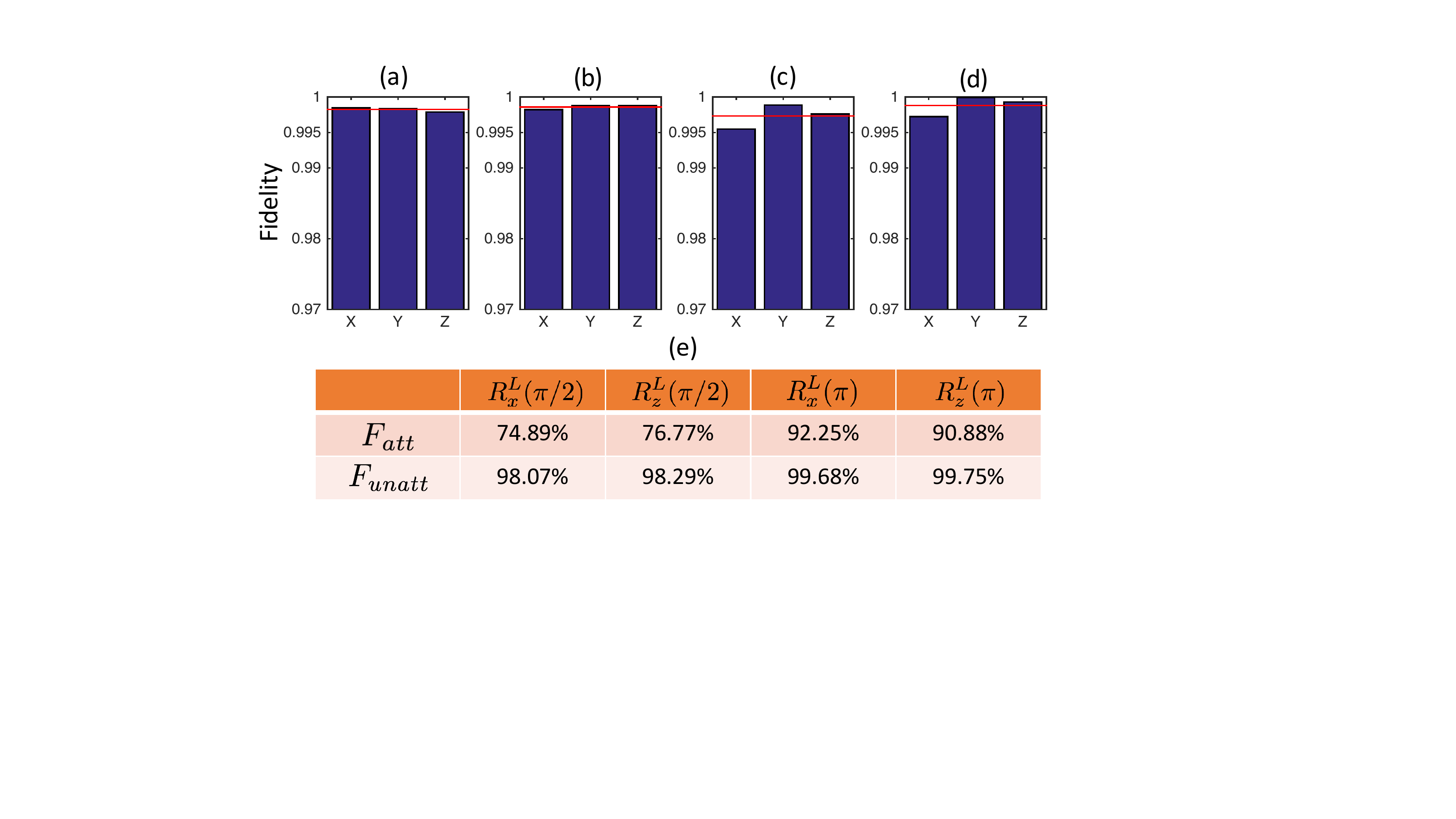}
\caption{The experimental unattenuated fidelities of different output states and table of experimental attenuated and unattenuated fidelities of matrix $\chi$ for the one-qubit NHQC gates. (a), (b), (c) and (d) give the fidelities of $\rho_{out}$ for $R_x^{L}(\pi/2)$, $R_z^{L}(\pi/2)$, $R_x^{L}(\pi)$ and $R_z^{L}(\pi)$ respectively, applied to input states $X$, $Y$ and $Z$. Here $X\equiv\sigma_x$, $Y\equiv\sigma_y$ and $Z\equiv\sigma_z$. The red solid horizontal lines are average fidelities of output states for their respective gates. (e) is a table of experimental attenuated and unattenuated fidelities of $\chi$ for $R_x^{L}(\pi/2)$, $R_z^{L}(\pi/2)$, $R_x^{L}(\pi)$ and $R_z^{L}(\pi)$.   }\label{f3}
\end{center}
\end{figure}

\section{Experimental results} 
\begin{figure}[!h]
\begin{center}
\includegraphics[width=8.5cm,angle=0]{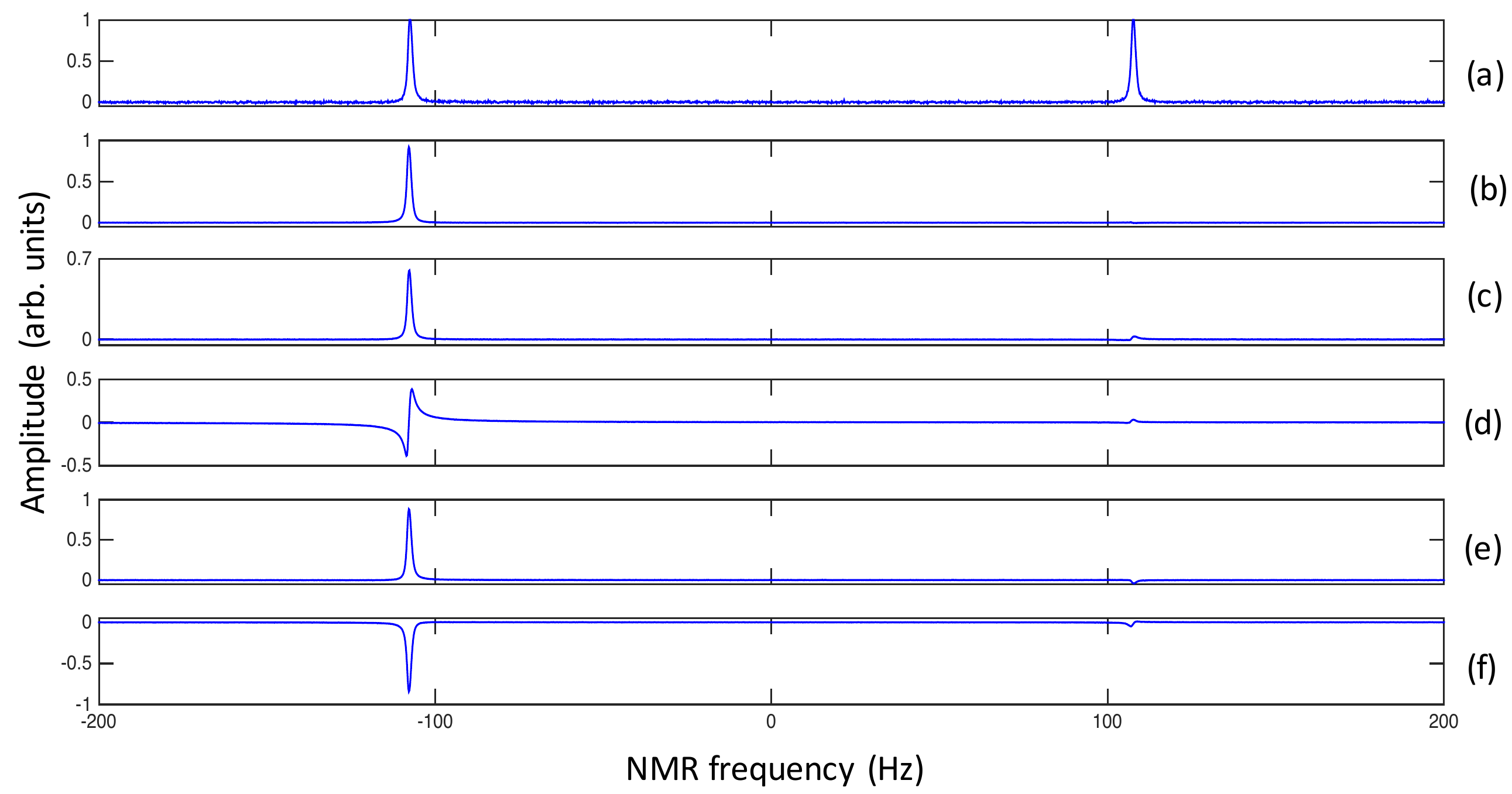}
\caption{Experimantal NMR spectra of $^{13}$C. (a) shows the spectrum of $^{13}$C when the system is in the thermal equilibrium state; (b) is the spectrum of input state where $\rho_{in}=\sigma_x$ and $^{1}$H is in state $|1\rangle\langle1|$; (c), (d), (e) and (f) are the spectra of output states after applying $R_x^{L}(\pi/2)$, $R_z^{L}(\pi/2)$, $R_x^{L}(\pi)$ and $R_z^{L}(\pi)$ to the input state $\sigma_x$, respectively.}\label{f4}
\end{center}
\end{figure}

The experimental output density matrix $\rho_{out}$ is determined by a procedure called quantum state tomography (QST)~\cite{Lee2002quantum,Wang2016}. To compare the experimental output density matrix $\rho_{out}$ and its theoretical matrix $\rho_{th}$, the attenuated and unattenuated state fidelities~\cite{fortunato2002design,weinstein2004quantum} $F_{att}(\rho)$ and $F_{unatt}(\rho)$ are calculated. Specifically, $F_{att}(\rho)$ and $F_{unatt}(\rho)$ are defined as $F_{att}(\rho)=\rm{Tr}\left(\rho_{out}\rho_{th}\right)/\sqrt{Tr\left(\rho_{th}\rho_{th}\right)Tr\left(\rho_{in}\rho_{in}\right)}$ and $F_{unatt}(\rho)=\rm{Tr}\left(\rho_{out}\rho_{th}\right)/\sqrt{Tr\left(\rho_{th}\rho_{th}\right)Tr\left(\rho_{out}\rho_{out}\right)}$, respectively. The average experimental attenuated state fidelities of output states of one-qubit NHQC gates $R_x^{L}(\pi/2)$, $R_z^{L}(\pi/2)$, $R_x^{L}(\pi)$ and $R_z^{L}(\pi)$ are $66.52\%$, $69.02\%$, $89.67\%$ and $87.84\%$, respectively, and their average experimental unattenuated fidelities are $99.82\%$, $99.86\%$, $99.73\%$ and $99.88\%$, respectively. The differences between the attenuated and unattenuated fidelities are mainly caused by signal loss. Fig.~\ref{f3} illustrates experimental unattenuated fidelities of output states for $R_x^{L}(\pi/2)$, $R_z^{L}(\pi/2)$, $R_x^{L}(\pi)$ and $R_z^{L}(\pi)$ applying to $\rho_{in}$ in Eq. (\ref{e5}), where the red solid lines are their individual average unattenuated fidelities in Fig.~\ref{f3}(a), (b), (c) and (d). Fig.~\ref{f4} illustrates some example experimental NMR spectra observed at $^{13}$C, in which (a) is the spectrum of the thermal equilibrium state; (b) is the spectrum of input state where $\rho_{in}=\sigma_x$ and $^{1}$H is in state $|1\rangle\langle1|$; (c), (d), (e) and (f) are the spectra of output states after applying $R_x^{L}(\pi/2)$, $R_z^{L}(\pi/2)$, $R_x^{L}(\pi)$ and $R_z^{L}(\pi)$ to the input state $\sigma_x$, respectively.

\begin{figure}
\begin{center}
\includegraphics[width=8.5cm,angle=0]{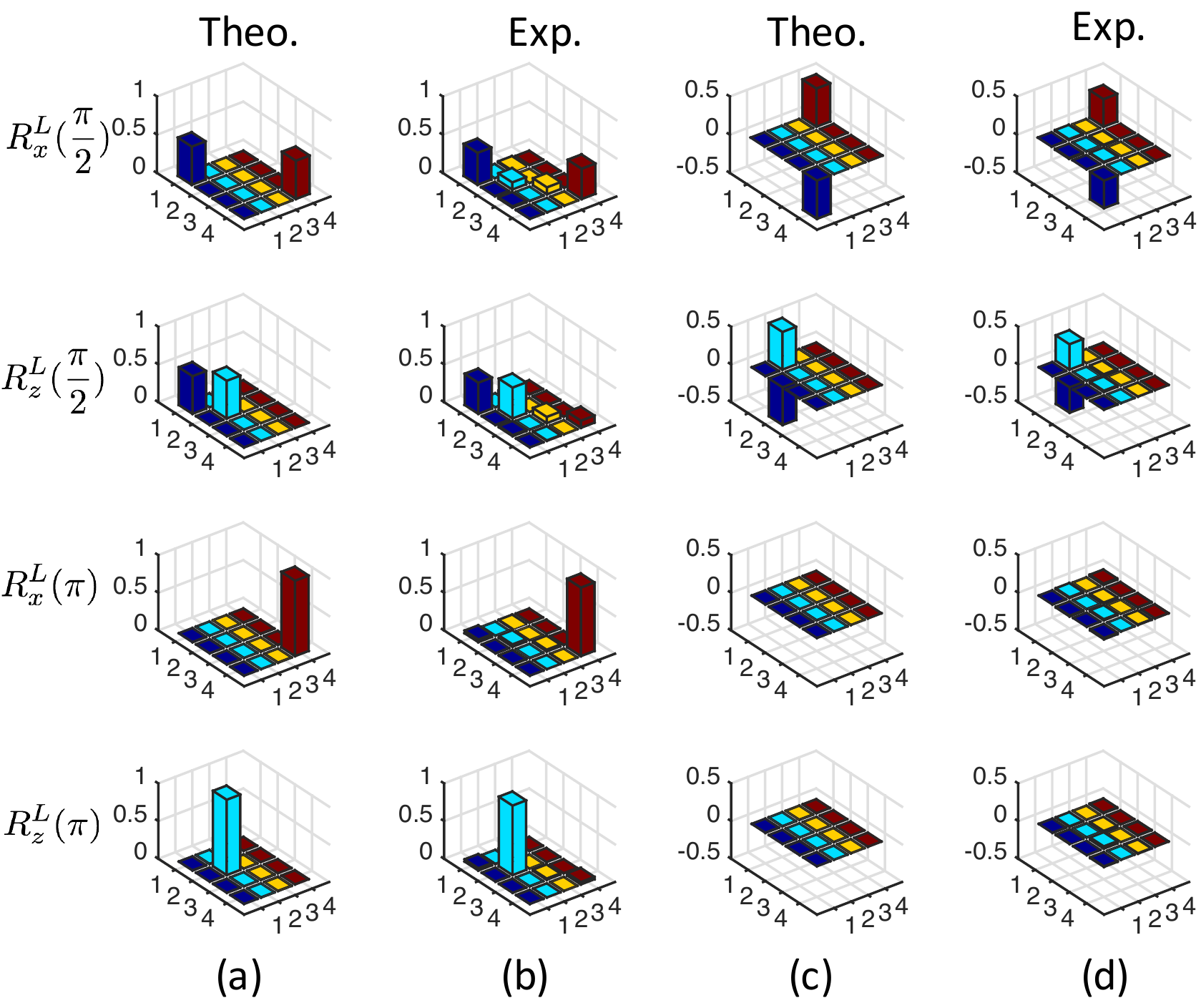}
\caption{The theoretical and experimental matrices $\chi$ for one-qubit NHQC gates $R_x^{L}(\pi/2)$, $R_z^{L}(\pi/2)$, $R_x^{L}(\pi)$ and $R_z^{L}(\pi)$. The (a) and (c) columns are the real and imaginary parts of theoretical $\chi$ matrices, respectively, while the (b) and (d) columns are the real and imaginary parts of experimental $\chi$ matrices,respectively. The numbers $1 - 4$ in the $x$ and $y$ axes represent the operators in the operator basis set $\{I, \sigma_x, -i\sigma_y, \sigma_z\}$.}\label{f5}
\end{center}
\end{figure}

As mentioned before, matrix $\chi$ can quantitatively characterize the quantum process $R_x^{L}(\pi/2)$, $R_z^{L}(\pi/2)$, $R_x^{L}(\pi)$ and $R_z^{L}(\pi)$ and we can prove a good implementation of one-qubit NHQC gates by determing experimental $\chi_{exp}$. We use the technique discribed in Ref.~\cite{nielsen2010quantum} to calculate the experimental QPT $\chi_{exp}$ matrix. Fig.~\ref{f5} shows the theoretical and experimental $\chi$ matrices for one-qubit NHQC gates, where the (a) and (c) columns are the real and imaginary parts of theoretical $\chi$ matrices, respectively, while the (b) and (d) columns are the real and imaginary parts of experimental $\chi$ matrices, respectively. Just like the fidelity of density matrix $\rho_{out}$, the attenuated and unattenuated fidelities between the theoretical $\chi_{th}$ and the experimental $\chi_{exp}$ are defined as $F_{att}(\chi)=|\rm{Tr}(\chi_{exp}\chi_{th}^{\dagger})|$ and $F_{unatt}(\chi)=|\rm{Tr}(\chi_{exp}\chi_{th}^{\dagger})|/\sqrt{Tr(\chi_{th}\chi_{th}^{\dagger})Tr(\chi_{exp}\chi_{exp}^{\dagger})}$~\cite{wang2008alternative,zhang2012experimental}. The experimental attenuated and unattenuated fidelities of matrices $\chi$ for $R_x^{L}(\pi/2)$, $R_z^{L}(\pi/2)$, $R_x^{L}(\pi)$ and $R_z^{L}(\pi)$ are presented in the table of Fig.~\ref{f4} (e). The difference between $F_{att}(\chi)$ and $F_{unatt}(\chi)$ are mainly caused by signal loss and $F_{unatt}(\chi)$ can represent the similarity between $\chi_{th}$ and  $\chi_{exp}$ ignoring certain errors due to signal loss.
\section{Summary} 
In this paper, we constructed two Hamiltonians for implementing single-shot NHQC gates $R_x^{L}(\pi/2)$, $R_z^{L}(\pi/2)$, $R_x^{L}(\pi)$ and $R_z^{L}(\pi)$. To implement $R_x^{L}(\pi/2)$ and $R_z^{L}(\pi/2)$ in NMR pulse sequences, the Trotter formula is employed, while $R_x^{L}(\pi)$ and $R_z^{L}(\pi)$ can be implemented directly in NMR pulse sequences without approximation. $^{13}$C labeled chloroform sample is used as our two-qubit quantum information processor, where $^{13}$C acts as the work qubit and $^{1}$H acts as the ancillary qubit. At first, we prepare the NMR system in $|1\rangle\langle1|\otimes\rho_{in}$ with a cat-state method; Then, at the stage of implementing the quantum circuits of Fig.~\ref{f2} (a), a sequence compiler developed for NMR quantum information processor based on Ref.~\cite{ryan2008liquid} is used to optimize the whole pulse sequence, minimizing the total error of the implementation compared to the intended  evolution; At last, a matrix $\chi$ is used to characterize the quantum process during conducting the QPT procedure to prove a good implementation of one-qubit NHQC gates. And the unattenuated fidelities of experimental $\chi$ for $R_x^{L}(\pi/2)$, $R_z^{L}(\pi/2)$, $R_x^{L}(\pi)$ and $R_z^{L}(\pi)$ are $98.07\%$, $98.29\%$, $99.68\%$ and $99.75\%$, respectively.To be noticed, the ancillary qubit stays in $|1\rangle\langle1|$ the whole time implementing the one-qubit NHQC gates. To our knowledge, this is the first experimental implementation of single-shot holonomic nonadiabatic holonomic gates reported in the literature. The methods and techniques used here can be extended to other quantum systems, like NV centers in diamond, superconducting qubits, trapped ions and so on.
\begin{acknowledgments}
The authors would like to thank D. Tong, G. Xu, S. Hou and G. Feng for helpful discussions, we also thank IQC, University of Waterloo, for providing the software package for NMR pulse sequence compiler. This work is supported by the National Natural Science Foundation of China under Grant No. 11175094, 91221205, and 11405093, and the National Basic Research Program of China under Grants No. 2015CB921002.
\end{acknowledgments}

\end{document}